\documentclass[prb,twocolumn,showpacs]{revtex4}

\usepackage{graphics}
\usepackage{amsmath}

\newcommand{\Journal}[4]{#1 \textbf{#2}, #3 (#4)}
\newcommand{\PRev}{Phys. Rev.}
\newcommand{\JPSJ}{J. Phys. Soc. Jpn.}
\newcommand{\be}{\begin{equation}}
\newcommand{\ee}{\end{equation}}
\newcommand{\nn}{\nonumber}
\newcommand{\eps}{\varepsilon}
\newcommand{\pr}{PrOs$_4$Sb$_{12}$}
\newcommand{\mb}[1]{\mathbf{#1}}

\begin{document}

\title{Superconducting states in the tetrahedral compound \pr}

\author{I. A. Sergienko}
\author{S. H. Curnoe}
\affiliation{Department of Physics and Physical Oceanography,
Memorial University of Newfoundland, St. John's, NL, A1B 3X7, Canada}

\begin{abstract}
We find possible superconducting states for tetrahedral ($T_h$) symmetry crystals 
with strong spin-orbit coupling using Landau theory. 
Additional symmetry breaking within the superconducting state is 
considered. We discuss nodes of the gap functions for the different states, secondary
superconducting order parameters and coupling to the elastic strain. By comparing our results
to experiments, we find that superconductivity in \pr\ is best described by the 
three-dimensional representations of point group $T_h$.
\end{abstract}


\maketitle

\section{Introduction}
The discovery of superconductivity in the heavy fermion compound 
\pr~\cite{Maple2001,Bauer2002} has
spawned a flurry of experimental~\cite{Kotegawa03,Aoki2002,Macl02,Oeschler2003,%
Vollmer2003,Izawa2003,Sugawara02,%
Kohgi03,Cao03,Tayama03,Ho03,Frederick03a,Aoki2003,Suderow03,Frederick03b,Chia03,Broun}
and theoretical~\cite{Sugawara02,Goryo2003,Maki2003,Miyake2003,Ichioka2003,Asano03,Anders2002} 
activity. \pr\ is the first Pr-based heavy fermion superconductor and the
first among the family of rare-earth filled skutterudite compounds.
The onset of superconductivity occurs at $T_{c1} = 1.85$ K;
an additional phase transition is observed as an anomaly in the
specific heat\cite{Vollmer2003} and
magnetisation \cite{Tayama03} at $T_{c2}=1.75$ K.
Thermal conductivity measurements in a rotating magnetic field
revealed the presence of nodes and a lowering of the symmetry of the gap 
function from four-fold (A-phase) to two-fold (B-phase) at 
$T_{c2}$.\cite{Izawa2003} Even more intriguing is the observation of 
broken time-reversal 
symmetry in the superconducting (SC) state.\cite{Aoki2003}  By all indications, \pr\ is a new kind
of unconventional superconductor.

A central issue in the study of unconventional superconductivity is the symmetry of the SC
order parameter. 
The phenomenological Landau theory approach is particularly useful when little is known about the
mechanism of superconductivity at a microscopic level, and is ideal for describing multiple 
phase transitions as is the case for \pr. The starting point is knowledge of the crystal
symmetry group, according to whose representations order parameters are classified.
The outcome of this approach is detailed knowledge of all possible phase diagrams and symmetry 
properties of the SC state, including nodes of the gap function.\cite{Volovik1985}
Phenomenological theory can also predict the order of the phase transition.
While the normal-to-SC phase transition is 
expected to be second order since third-order terms in the Landau potential 
expansion\cite{Landau68} are prohibited because of gauge symmetry, 
this is not generally the case for 
phase transitions within the SC state. 

Several theoretical models of the SC order parameter in \pr\ have been proposed
in order to account for the experimental data. Goryo suggested different combinations of 
$s$- and $d$-wave gap functions for the A- and B-phases~\cite{Goryo2003}
in order to account for the change in symmetry observed in the thermal conductivity
experiment.\cite{Izawa2003} The A-phase was assumed to have an anisotropic $s$-wave gap function 
that has six minima along the 
$[100]$, $[010]$, and $[001]$ directions. In the B-phase, an $s+i d_{z^2-x^2}$-wave 
combination was proposed. Different $s+g$-wave basis functions were proposed by 
Maki {\em et al.} for both states.\cite{Maki2003}
An  $f$-wave pairing state with weak spin-orbit coupling
was proposed by Ichioka {\em et al.} to describe a state with point nodes on all three 
axes.\cite{Ichioka2003} Finally, Miyake {\em et al.} considered a microscopic model based on
quadrupolar fluctuations and nesting in the Fermi surface, and argued in favour of 
$p_x+i p_y$-wave pairing.\cite{Miyake2003}

While the models mentioned above
may describe particular experiments, they can 
only be considered as empirical. There are at least two fundamental shortcomings:
(i) the models are in fact based on the assumption that the point group crystal symmetry is 
$O_h$. \pr\ has lower $T_h$ symmetry (space group $Im{\bar 3}$, $T_h^5$); (ii) there is no 
physical reason why the system should 
choose one particular combination of the basis functions of the irreducible representation
of the symmetry group over the others. Strictly speaking, the theory allows
all basis functions to contribute to the gap function. Moreover, the 
coefficients in the such combinations in general depend on the external conditions (temperature,
magnetic field, \emph{etc.}). Only such a general
state is thermodynamically stable and
occupies a finite region of the phase diagram. 

In this paper we use the Landau theory approach to classify SC phases for tetrahedral ($T_h$) 
crystals, including those which may be reached by additional symmetry breaking within the SC 
state. We use the strong spin-orbit coupling limit in which the spin rotation symmetry is
broken.\cite{Sigrist1991,HarimaNote} 

The first attempt to accomplish such a classification was made by Gufan.\cite{Gufan1995} 
In Sec.\ II of this paper we use a different approach and reproduce most results of 
Ref.~\onlinecite{Gufan1995} for $T_h$ symmetry.\cite{mistake} In addition, we 
discuss the basis functions of the irreducible representations, the 
gap function nodes and the orders of the phase transitions 
between different SC states.  In Sec.\ III we consider 
secondary SC order parameters which
influence the nodes of the gap functions. In Sec.\ IV the coupling between
the SC order parameters and elastic strain is discussed.  Sec.\ V is
devoted to matching the experimental data with the states found theoretically.
Sec.\ VI summarizes the paper.

\section{Classification of superconducting states}

A procedure for constructing SC classes and finding the gap nodes with
strong spin-orbit coupling was originally proposed by Volovik and Gor'kov (VG),\cite{Volovik1985}
who listed all SC states which can be reached from the normal state by a second
order phase transition for $O_h$, $D_{4h}$ and $D_{6h}$ crystals. 
One begins by classifying possible order parameters according
to the representations of the crystal point group.  In systems with
inversion symmetry, all representations have a definite parity.  Those
with even parity must be matched with singlet pairing of the spin
states for the pair wavefunction to be antisymmetric; likewise
odd parity representations are matched with triplet spin states.
For each parity, the group $T_h$ has a one dimensional representation $A$, a two dimensional
representation $E$, which is reducible to two one-dimensional representations 
that are complex conjugate, and a three dimensional representation $T$.\cite{Tinkham}

The SC gap function is a $2 \times 2$ matrix in pseudospin space given 
by $\widehat \Delta(\mathbf{k})=i\widehat\sigma_y \psi(\mathbf{k})$ for singlet pairing and by 
$\widehat \Delta(\mathbf{k})=i(\mathbf{d}(\mathbf{k})\boldsymbol{\widehat\sigma})\widehat
\sigma_y$ for triplet pairing, where $\boldsymbol{\widehat\sigma}=(\widehat\sigma_x, \widehat
\sigma_y, \widehat\sigma_z)$ are Pauli matrices, $\psi(\mathbf{k})$ is an 
even scalar function and $\mathbf{d}(\mathbf{k})$ is
an odd pseudovector function. $\psi(\mathbf{k})$ and $\mathbf{d}(\mathbf{k})$
are expressed in terms of the components of the order parameter
$\eta_i$ as $\psi(\mathbf{k})=\sum_i \eta_i \psi_i(\mathbf{k})$ and
$\mathbf{d}(\mathbf{k})=\sum_i \eta_i \mathbf{d}_i(\mathbf{k})$.
Here $\psi_i(\mathbf{k})$ and $\mathbf{d}_i(\mathbf{k})$ are the basis functions for 
the even (spin-singlet case) and odd (spin-triplet case) irreducible representations of the point 
group, respectively.\cite{Sigrist1991}

The method of finding the SC states implemented by VG is to construct a Landau energy 
functional of $\eta_i$ for each order parameter that is invariant under 
$G\times U\times {\cal K}$, where 
$G$ is the point group, $U$ is gauge symmetry and ${\cal K}$ is time-reversal, and analyse
its extrema. In order to account for all possible phase diagrams, a
very large number of terms must be included and the analysis of such
a cumbersome model is tedious at best.\cite{TediousNote}
In practice, terms are restricted to those needed to describe the normal to superconducting phase
transition,\cite{Volovik1985,Blount1985} while states resulting from additional phase 
transitions within the SC state must be found by other methods.
The VG approach can be applied $T_h$ crystals. However, here we use even an easier method.
We use the fact that $T_h$ is a subgroup of $O_h$. Therefore we start with the results for
the symmetry groups of SC classes obtained by VG for $O_h$ symmetry and \emph{reduce} them by 
removing the symmetry elements that are absent in the normal state of $T_h$ symmetry. 

We consider additional symmetry breaking within the SC state by constructing
\emph{effective} Landau functionals of \emph{effective} order parameters, 
which describe the phase transitions
between SC states with a group-subgroup relation. 
This procedure is straightforward, since the symmetry
group of a SC state is discrete (the continuous gauge symmetry is 
already broken). In the following,
we consider the two dimensional representation in detail, 
while only the results
are given for the three dimensional representation.

\begin{table*}[t]
\caption{\label{tbl} SC states described by one irreducible representation of the
point group $T_h$.
The relative magnitudes and phases of the
components of the order parameter are defined in the first column. The
symmetry groups of the SC states are listed in the second column.
Approximate and rigorous nodes of the gap function for even parity are listed in the third and 
fourth columns, similarly for odd parity in the fifth and sixth columns.
The word `same' is used  when rigorous nodes coincide with approximate nodes.
The numbers in parentheses in the fifth column indicate whether one or both gaps have
nodes.}
\begin{ruledtabular}
\begin{tabular}{clllllll}
 State & Symmetry & & Approximate & Rigorous & & Approximate & Rigorous\\
       &          & & nodes       & nodes    & & nodes & nodes\\
\hline
(1) & $T\times{\cal K}$ & $A_g$ & none & none & $A_u$ & none & none\\ \hline
(1, 0) & $T(D_2)$ & &8 points $\langle 111\rangle$ & same & &8 points $\langle 111\rangle$ (1)
& same\\ 
($\phi_1$, $\phi_2$) & $D_2 \times {\cal K}$ & $E_g$ & 8 points $\langle 111\rangle$ 
& none &$E_u$ & none & none\\
($\eta_1$, $\eta_2$) & $D_2$ & & 8 points $\langle 111\rangle$ & none & &none
& none\\
 \hline
(1, 0, 0) & $D_2(C_2)\times {\cal K}$ & & 2 lines $k_y=0$, $k_z=0$ & same & & 2 points [100](2)
& same \\
(1, 1, 1) & $C_3 \times {\cal K}$ & & 6 points $\langle 001\rangle$ & none && none & none\\
(1, $\eps$, $\eps^2$) & $C_3 (E)$ & & 6 points $\langle 001\rangle$, 2 points [111]
& 2 points [111]&  & 2 points [111](1) & same\\
($|\eta_1|, i|\eta_2|,0$) & $D_2(E)$ & & 1 line $k_z=0$, 2 points [001] & same && none & none\\
$(|\eta_1|, |\eta_2|, 0)$
& $C_2(E)\times {\cal K}$ & $T_g$ & 1 line $k_z=0$, 2 points [001] & same & $T_u$
& none & none\\
$(\eta_1, \eta_2, 0)$
& $C_2(E)$ & & 1 line $k_z=0$, 2 points [001] & same & & none & none\\
$(|\eta_1|, i|\eta_2|, |\eta_3|)$
& $C'_2(E)$ & & 6 points $\langle 001\rangle$ &  none & & none & none\\
$(|\eta_1|, |\eta_2|, |\eta_3|)$ & ${\cal K}$ &
& 6 points $\langle 001\rangle$ & none & & none & none\\
$(\eta_1, \eta_2, \eta_3)$
& $E$ & & 6 points $\langle 001\rangle$ & none & & none & none\\
\end{tabular}
\end{ruledtabular}
\end{table*}

Our results are summarized in Table~\ref{tbl}. It lists all possible SC states for both even 
parity and odd parity which are possible if only a single irreducible 
representation is present.  
We define the relations between the components of the order parameters, the symmetry of the SC 
state, and the structure of nodes in the gap function. We make the
distinction between accidental, approximate
and rigorous nodes. \emph{Accidental} nodes occur in empirical models when a particular
form of the gap function is chosen \emph{a priori}, such as that
proposed in Ref.~\onlinecite{Maki2003}.  Such nodes cannot be stable because even small
contributions of functions with the same symmetry remove them immediately.\cite{Yip93}
Accidental nodes are unphysical and so we disregard them. \emph{Approximate} nodes are
a property of all possible basis functions which can be constructed for a given representation.
These nodes may be removed when admixtures of other representations,
which couple to the SC state as secondary order parameters,
are taken into account, thus leaving only \emph{rigorous} nodes required by the symmetry of the
SC state.\cite{Volovik1985,Monien1986}
The secondary order parameters are proportional to the third 
power of the primary order parameter.\cite{Monien1986}
Hence, the experiments that probe the symmetry of the gap function
close to $T_c$ may find the approximate nodes, while only the rigorous nodes remain when 
$T \rightarrow 0$.
A more detailed discussion of the secondary order parameters is given in
Sec.\ III.

\subsection{1D representation $A_{g,u}$}
The analysis of the one-dimensional representations $A_g$ and $A_u$ is
straightforward. Only gauge symmetry is broken and there are no nodes.
The symmetry of the SC state is $T\times {\cal K}$.  
In the lowest order in $\mb k$, the basis function for the singlet channel $\psi(\mathbf{k})$ is
constant on the Fermi surface and for the triplet channel 
$\mb{d}(\mb k)\sim k_x \mathbf{\hat x}
+k_y \mathbf{\hat y} + k_z \mathbf{\hat z}$.
Here and below `$\sim$' means `transforms like' so that all our results remain valid
for higher order basis functions.

\subsection{2D representation $E_{g,u}$}
We choose the basis functions of the two-dimensional representations $E_g$ and $E_u$
in complex form as in Ref.~\onlinecite{Volovik1985},
\be
\begin{array}{ll}\label{basEg}
\psi_1\sim k_x^2+\eps k_y^2+\eps^2 k_z^2, & \psi_2\sim k_x^2+\eps^2 k_y^2+\eps k_z^2;\\
\mathbf{d_1}\sim k_x \mathbf{\hat x} +\eps k_y\mathbf{\hat y}+\eps^2 k_z\mathbf{\hat z}, &
\mathbf{d_2}\sim k_x \mathbf{\hat x} +\eps^2 k_y\mathbf{\hat y}+\eps k_z\mathbf{\hat z},
\end{array}
\ee
where $\eps=\exp (2\pi i/3)$. With this choice of the basis functions, the order parameter 
has the following transformation properties,
\begin{eqnarray}
C_2 (\eta_1, \eta_2) &=& (\eta_1, \eta_2),\nn\\
C_3^{111} (\eta_1, \eta_2) &=& (\eps\eta_1, \eps^2\eta_2),\nn\\
{\cal K} (\eta_1, \eta_2) &=& (\eta_2^*, \eta_1^*),\nn\\
U(\theta) (\eta_1, \eta_2) &=& e^{i\theta}(\eta_1, \eta_2), \label{2Delements}
\end{eqnarray}
where $C_2$ stands for any of the twofold rotations in $T_h$, and $C_3^{111}$ is a $2\pi/3$ 
rotation about the $[111]$ direction, and $U(\theta)$ is a gauge transformation.

In Table I, three states are listed for the two-dimensional representations
of $T_h$. 
These differ from the $O_h$  states
$(1,0)$, $(1,1)$ and $(1,-1)$.  
As shown below, the extra freedom in
the phase and magnitude of the last two states of $T_h$ arise from terms in the
free energy which are allowed under $T_h$ but not $O_h$. 

The SC state $(1, 0)$ in $O_h$ corresponds to the group\cite{Volovik1985}
\begin{eqnarray}
O(D_2)& =& \{D_2, 2C_4^x{\cal K},2C_4^y U(2\pi/3){\cal K},
 2C_4^zU(4\pi/3){\cal K}, \nonumber \\
& & 2C_2^{yz}{\cal K},2C_2^{xz}U(2\pi/3){\cal K},
2C_2^{xy}U(4\pi/3){\cal K}, \nonumber \\
 & & 4C_3 U(4\pi/3),4C_3^2U(2\pi/3)\},
\end{eqnarray}
where $D_2$ is the group of two-fold rotations about the [100], [010], and [001] axes.
In $T_h$, the remaining symmetry elements are
\be
T(D_2)=\{D_2, 4 U(4\pi/3)C_3, 4 U(2\pi/3)C_3^2\}.
\ee

Considering the
symmetry groups of the states $(1, 1)$ and $(1,-1)$ in $O_h$,
which are $D_4\times {\cal K}$ and $D_4(D_2)\times {\cal K}$ respectively,\cite{Volovik1985}
where
\be
D_4(D_2)\times {\cal K} = \{D_2,2C_4^xU(\pi),2C_2^{yz}U(\pi) \}\times{\cal K},
\ee
we notice that they both reduce to the same symmetry $D_2\times \cal K$ in $T_h$.
Moreover, it follows from~(\ref{2Delements}) that this symmetry does not fix the relation between
the phases $\phi_1$ and $\phi_2$ of the OP components $\eta_{1,2}=|\eta_{1,2}|\exp(i\phi_{1,2})$,
but the magnitudes are equal $|\eta_1|=|\eta_2|$. 
Therefore, we denote this state as $(\phi_1,\phi_2)$. This  also
may be verified from the following Landau model which describes the $E_{g,u}$ 
representation of $T_h$:
\begin{eqnarray}
F&=&\alpha(|\eta_1|^2+|\eta_2|^2) + \beta_1 (|\eta_1|^4+|\eta_2|^4) +
2\beta_2 |\eta_1|^2|\eta_2|^2\nn\\
&& + \gamma_1 (\eta_1^3\eta_2^{*3}+\eta_2^3\eta_1^{*3})
+ \gamma_2 i(\eta_1^3\eta_2^{*3}-\eta_2^3\eta_1^{*3}),
\label{F_Eg}
\end{eqnarray}
where $\alpha$, $\beta_1$, $\beta_2$, $\gamma_1$, and $\gamma_2$ are phenomenological parameters.
The last two terms reduce to $2\gamma_1 |\eta_1|^3|\eta_2|^3 \cos(3\phi)
+ 2\gamma_2 |\eta_1|^3|\eta_2|^3 \sin(3 \phi)$, where $\phi \equiv \phi_1-\phi_2$. 
Thus the equilibrium value of $\phi$ depends on the (generally temperature dependent) ratio 
$\gamma_1/\gamma_2$. In 
contrast, in $O_h$ symmetry the $\gamma_2$ term is prohibited, hence $\phi$ is fixed
to be either $0$ [for $(1,1)$ state] or $\pi$ [for $(1,-1)$ state]. 

The gap function of the $(\phi_1, \phi_2)$ state in the singlet channel is\cite{ComNodes}
\begin{eqnarray}
\Delta(\mathbf{k})
&\sim & \cos (\phi/2) k_x^2+\cos(\phi/2+2\pi/3) k_y^2\nn\\
&& + \cos(\phi/2+4\pi/3) k_z^{2}, \label{gapEg}
\end{eqnarray}
and in the triplet channel,
\begin{eqnarray}
\Delta(\mathbf{k})
&\sim & \cos^2 (\phi/2) k_x^2+\cos^2(\phi/2+2\pi/3) k_y^2\nn\\
&& + \cos^2(\phi/2+4\pi/3) k_z^{2}.
\label{gapEg_tr}
\end{eqnarray}
We would like to stress that the state $d_{x^2-y^2}$ and its equivalents, obtained by permutations 
of $x$, $y$, and $z$, are \emph{not} stable in $T_h$. Instead, they are replaced by the more 
general state $(\phi_1, \phi_2)$ with the gap function~(\ref{gapEg}).

In $O_h$, the states $(1, 0)$, $(1, 1)$, and $(1, -1)$ are connected to the normal state by a 
second order phase transition.\cite{Volovik1985} 
Since up to fourth order terms, the model~(\ref{F_Eg}) coincides 
with that of $O_h$, we conclude that the states $(1, 0)$ and $(\phi_1,\phi_2)$ can be 
reached from the normal state in $T_h$ by a second order phase transition.

There is a third state which can be described by the $E$ representation in $T_h$. Its symmetry group 
is $D_2$ (time reversal is broken), which is a common subgroup of both
$T(D_2)$ and $D_2 \times {\cal K}$. As is seen 
from~(\ref{2Delements}), 
it has no constraints on either the magnitudes or phases, therefore
we denote this state $(\eta_1,\eta_2)$.

The phase transitions $(1,0)\rightarrow (\eta_1,\eta_2)$ and
$(\phi_1,\phi_2)\rightarrow (\eta_1,\eta_2)$ may be described
by {\em effective} free energy functionals of effective order parameters.
The factor group $T(D_2)/D_2$ is isomorphic to the
cyclic group $C_3$ and thus third order
terms are allowed in the effective free energy describing the
$(1,0)\rightarrow (\eta_1,\eta_2)$ transition. This implies that it cannot be
second order. The effective free energy is a functional of the effective order 
parameter $\eta_2$,
\begin{eqnarray}
F_{eff}[(1,0)\rightarrow (\eta_1,\eta_2)] & = & \tilde\alpha|\eta_2|^2
+\tilde\gamma_1(\eta_2^3+\eta_2^{*3})\nonumber \\
& & +i\tilde\gamma_2(\eta_2^3-\eta_2^{*3}) + \tilde\beta|\eta_2|^4
\label{2Deff1}
\end{eqnarray}
On the other hand, a second order transition
$(\phi_1,\phi_2)\rightarrow(\eta_1,\eta_2)$ is possible.
This transition is described by an effective order parameter $\delta \equiv |\eta_1|-|\eta_2|$.  
Third order terms are prohibited in the effective free energy because of time reversal symmetry,
\be
F_{eff}[(\phi_1,\phi_2) \rightarrow (\eta_1,\eta_2)] = \alpha' \delta^2 + \beta' \delta^4.
\label{2Deff2}
\ee

There are no other states described by the $E$ representation alone,
because the basis functions~(\ref{basEg}) are invariant with respect to all symmetry operations
of $D_2$ group and there are no other symmetry groups containing $D_2$.

\begin{figure}
\caption{\label{fig1}Second order phase transitions among states of
the $E_g$ and $E_u$ representations of $T_h$.}
\includegraphics{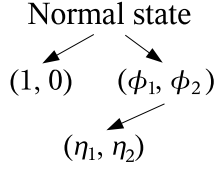}
\end{figure}

\subsection{3D representation $T_{g,u}$}

The lowest order basis functions for the $T_g$ representation of $T_h$ are
`$d$-wave' ({\em i.e.} second order in $\mb k$), 
\be
\psi_1 \sim k_y k_z, \qquad \psi_2 \sim k_x k_z, \qquad \psi_3 \sim k_x k_y,
\ee
while for the $T_u$ representation the lowest order basis functions are
`$p$-wave', and there are two independent sets of them,
\begin{eqnarray}
\mathbf{d_1} &\sim& a k_y \mathbf{\hat z} + b k_z \mathbf{\hat y},\nn\\
\mathbf{d_2} &\sim& a k_z \mathbf{\hat x} + b k_x \mathbf{\hat z},\nn\\
\mathbf{d_3} &\sim& a k_x \mathbf{\hat y} + b k_y \mathbf{\hat x}.
\end{eqnarray}
Here $a$ and $b$ are arbitrary numbers, in contrast to $O_h$, which fixes
$b=-a$ in
$T_{1u}$ representation and $b=a$ in $T_{2u}$ representation.
It follows that the order parameter transforms as
\begin{eqnarray}
C_2^z (\eta_1,\eta_2,\eta_3) &=& (-\eta_1,-\eta_2,\eta_3), \nn\\
C_3^{111} (\eta_1,\eta_2,\eta_3) &=& (\eta_2,\eta_3,\eta_1), \nn\\
{\cal K} (\eta_1,\eta_2,\eta_3) &=& (\eta_1^*,\eta_2^*,\eta_3^*),\nn\\
U(\theta) (\eta_1, \eta_2, \eta_3) &=& e^{i\theta}(\eta_1, \eta_2, \eta_3). \label{3Delements}
\end{eqnarray}

To find the SC states of the three dimensional representation,
we again use the $O_h$ states as a starting point.
For $O_h$, there are four states accessible by a second order phase transition from the
normal state: $(1,0,0)$, $(1,i,0)$, $(1,1,1)$ and $(1,\eps,\eps^2)$,
with symmetries 
\begin{eqnarray}
D_4(C_4)\times {\cal K} &=& \{E, C_2^x, 2C_4^x, 4U(\pi)C_2^{\perp x}\}\times {\cal K},\nn\\ 
D_4(E) &=& \{E, U(\pi)C_2^x, 2U(\pm\pi/2)C_4^x, C_2^x {\cal K}, \nn\\
&& \quad U(\pi)C_2^y{\cal K}, 2U(\pm\pi/ 2)C_2^{xy}{\cal K}\},\nn\\
D_3(C_3)\times {\cal K} &=& \{E, 2C_3, 3 U(\pi) C_2^{xy} \}\times {\cal K}\nn\\
D_3(E) &=& \{ E, U(4\pi/3)C_3, U(2\pi/3)C_3^2, C_2^{yz}{\cal K},  \nn\\
&& U(2\pi/3) C_2^{xz}{\cal K}, U(4\pi/3) C_2^{xy}{\cal K}\},
\end{eqnarray}
respectively.\cite{Volovik1985} Here $E$ is the identity element.
Reducing these groups, we find the following classes for $T_h$:
\begin{eqnarray}
D_2(C_2)\times {\cal K}&=& \{E, C_2^x, U(\pi) C_2^y, U(\pi) C_2^z\}\times {\cal K}\nn \\
D_2(E) & =& \{E, U(\pi) C_2^z, C_2^x {\cal K}, U(\pi)C_2^y{\cal K}\} \nn \\
C_3 \times {\cal K}&=& \{E, C_3, C_3^2\}\times {\cal K}\nn \\
C_3(E)&=& \{ E, U(4\pi/3)C_3, U(2\pi/3)C_3^2\}
\label{Tclasses}
\end{eqnarray}
We notice that the 
$D_2(E)$ symmetry actually does not require $|\eta_1| = |\eta_2|$. Hence, the state
$(1, i, 0)$ is not stable in $T_h$. Instead, it is replaced by the state 
$(|\eta_1|,i|\eta_2|,0)$. A direct second order normal-to-$(|\eta_1|,i|\eta_2|,0)$ transition is 
possible in 
$T_h$. These findings are also evident in the form of the Landau potential for the 3D order 
parameter. In order to display the $T_h$ (but not $O_h$) symmetry, a Landau model for $T_g$ and 
$T_u$ must include at least sixth order terms, as in the case of $E_g$ and $E_u$.
These sixth order terms are composed of five 
linearly independent invariants:
\be
\begin{array}{l}
|\eta_1|^6+|\eta_2|^6+|\eta_3|^6, \qquad |\eta_1|^2|\eta_2|^2|\eta_3|^2,\\
(|\eta_1|^2+|\eta_2|^2+|\eta_3|^2)(\eta_1^2\eta_2^{*2}+\eta_2^2\eta_3^{*2}+\eta_3^2\eta_1^{*2}
+ \text{c.c.}),\\
(|\eta_1|^4|\eta_2|^2+|\eta_2|^4|\eta_3|^2+|\eta_3|^4|\eta_1|^2)\\
\pm (|\eta_1|^2|\eta_2|^4+|\eta_2|^2|\eta_3|^4+|\eta_3|^2|\eta_1|^4),\\
(\eta_1^4\eta_2^{*2}+\eta_2^4\eta_3^{*2}+\eta_3^4\eta_1^{*2})
\pm
(\eta_2^{4}\eta_1^{*2}+\eta_3^{4}\eta_2^{*2}+\eta_1^{4}\eta_3^{*2})+ \text{c.c.}. 
\label{6od_terms}
\end{array}
\ee
The negative signs in the last two invariants in~(\ref{6od_terms}) occur in
$T_h$ but not in $O_h$. 

\begin{figure}
\caption{\label{fig2}Second order phase transitions among states of
the $T_g$ and $T_u$ representations of $T_h$.}
\includegraphics{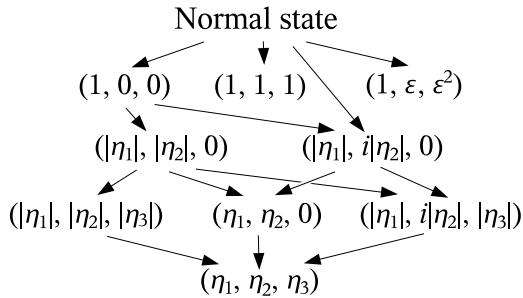}
\end{figure}

Considering all possible subgroups of the groups
in~(\ref{Tclasses}), we find five more SC states as
listed in Table~\ref{tbl}, where
\begin{eqnarray}
C_2(E) &=& \{E, U_1(\pi)C_2^z\}\nn,\\
C'_2(E) &=& \{E, U_1(\pi)C_2^y {\cal K}\}.
\end{eqnarray}
We have examined the transitions within the SC state by considering effective free
energies which describe them, similar to those described for
the 2D order parameter, Eqs.~(\ref{2Deff1}), (\ref{2Deff2}).  The diagram of all
second order phase transitions described by the three-dimensional representations of $T_h$ is given
in Fig.~\ref{fig2}.

Thus we find that the absence of four-fold rotation symmetry in \pr\ essentially
changes the structure of possible SC states. The states $(1,1)$ and $(1,-1)$
are not stable, because the value of $\phi$ in
Eq.~(\ref{gapEg}) is not fixed. Similarly,
the state $(1,i,0)$ is absent
in the three-dimensional representations.
Additionally, all SC states which may be connected to the normal state in $O_h$, $D_{4h}$ or 
$D_{6h}$ symmetry by a second order phase transition are \emph{one-parameter} in the sense that all
components of the order parameter are proportional to one quantity,  its absolute
value.\cite{Volovik1985, Blount1985}
The situation is different for the states
$(\phi_1, \phi_2)$ and $(|\eta_1|, i|\eta_2|, 0)$ in
$T_h$, for which two independent quantities describe the SC state.

\section{Secondary SC Order Parameters}

In general, the primary order parameter is accompanied by 
secondary order
parameters which do not change the symmetry of the SC state.
The influence of secondary order parameters on the gap nodes was discussed in
Sec.\ I.   
Since secondary order parameters do not change the overall symmetry of the
superconducting state, they are most easily found by identifying
supergroups of the states listed in the second column of Table~\ref{tbl} which
correspond to another superconducting state.  Table~\ref{tbl2} lists them.

In order to calculate how the secondary order parameters appear in the ordered phases, we need 
invariants of the types $\eta^3\xi$ and $\eta^{2}\xi^2$, where $\eta$ is the primary order 
parameter and $\xi$ is the secondary order parameter. 
From the first type of invariant, it is clear that 
$\xi$ and $\eta$ must have the same parity.
There are three scenarios to consider: i) The 2D primary order parameter with
1D secondary OP ii) 3D primary with 1D secondary and iii) 3D primary with
2D secondary. In the rest of this section, we denote the primary order parameters as
$\eta_j=|\eta_j|e^{i\phi_j}$ and the secondary order parameters as $\xi_j=|\xi_j|e^{i\theta_j}$.

\subsection{2D primary with 1D secondary}
The coupling terms of the two order parameters in the Landau potential are
\be
\begin{array}{l}
(\eta_1\eta_2^{*2}\xi+\eta_1^2\eta_2^*\xi^*)+(\eta_1^{*2}\eta_2\xi
+\eta_1^*\eta_2^2\xi^*),\\
i[(\eta_1\eta_2^{*2}\xi+\eta_1^2\eta_2^*\xi^*)-(\eta_1^{*2}\eta_2\xi
+\eta_1^*\eta_2^2\xi^*)],\\
\eta_1^*\eta_2^*\xi^2+\eta_1\eta_2\xi^{*2}, \quad
(|\eta_1|^2+|\eta_2|^2)|\xi|^2 .\\
\end{array}
\ee
In the state $(1, 0)$, 
the first two terms vanish, hence $\xi=0$. In the 
state $(\phi_1,\phi_2)$, the first two terms are finite and $|\xi| \propto |\eta|^3$. Minimization
with respect to 
$\theta$ yields $\theta=\frac{1}{2}(\phi_1+\phi_2)$. This relation between the phases
of the OP's ensures that
time reversal symmetry in preserved. There is no such relation between the 
phases when the primary order parameter state is $(\eta_1,\eta_2)$. This reflects the fact that 
time reversal symmetry is broken.

\subsection{3D primary and 1D secondary}
The coupling terms are
\be
\begin{array}{l}
(\eta_1\eta_2^*\eta_3^*+\eta_1^*\eta_2\eta_3^*+\eta_1^*\eta_2^*\eta_3)\xi+
\text{c.c},\\
(\eta_1^{*2}+\eta_2^{*2}+\eta_3^{*2})\xi^2+\text{c.c},\\
(|\eta_1|^{2}+|\eta_2|^{2}+|\eta_3|^{2})|\xi|^2
\end{array}
\ee
It follows that if any of the components of the 3D order parameter is
zero then the potential
has a  minimum at $\xi=0$. This is also the case for the state $(1, \eps, \eps^2)$. 
In the states in which $\phi_1=\phi_2=\phi_3$ [\emph{i.\ e.\ } $(1, 1, 1)$
and $(|\eta_1|, |\eta_2|, |\eta_3|)$], one obtains $\theta=\phi_1$.
However, in the state $(|\eta_1|, i|\eta_2|, |\eta_3|)$ we find $\theta=\phi_1\pm\frac{\pi}{2}$. 

\subsection{3D primary and 2D secondary}

The coupling terms are
%\begin{widetext}
\be
\begin{array}{l}
(\eta_1^*\eta_2\eta_3+\eps\eta_1\eta_2^*\eta_3+\eps^2\eta_1\eta_2\eta_3^*)
\xi_1^* \\
\quad +(\eta_1\eta_2^*\eta_3^*+\eps\eta_1^*\eta_2\eta_3^*+\eps^2\eta_1^*\eta_2^*
\eta_3)\xi_2+\text{c.c},\\
i[(\eta_1^*\eta_2\eta_3+\eps\eta_1\eta_2^*\eta_3+\eps^2\eta_1\eta_2\eta_3^*)
\xi_1^* \\ \quad
-(\eta_1\eta_2^*\eta_3^*+\eps\eta_1^*\eta_2\eta_3^*+\eps^2\eta_1^*\eta_2^*
\eta_3)\xi_2-\text{c.c}],\\
(\eta_1^{*2}+\eta_2^{*2}+\eta_3^{*2})\xi_1^*\xi_2^*+\text{c.c},\\
(|\eta_1|^2+\eps|\eta_2|^2+\eps^2|\eta_3|^2)\xi_1\xi_2^*+\text{c.c},\\
(|\eta_1|^2+|\eta_2|^2+|\eta_3|^2)(|\xi_1|^2+|\xi_2|^2).
\end{array}
\ee
%\end{widetext}

For this type of mixing we only consider the $(1, \eps, \eps^2)$ state of
the primary order parameter, since in the other states where $E_{g,u}$ is present as a
secondary order parameter, $A_{g,u}$ is also present, and it surely
removes all nodes. The first two invariants in the $(1, \eps, \eps^2)$ state
reduce to $6|\eta_1|^3|\xi_2|\cos(\theta_2-\phi_1)$ and
$6|\eta_1|^3|\xi_2|\sin(\theta_2-\phi_1)$, respectively. Thus, the state
$(0, 1)$, which is equivalent to $(1,0)$, appears as a secondary effect. Note
that $\theta_2-\phi_1$ is not fixed, which is expected since the state breaks time
reversal symmetry.

\begin{table}[t]
\caption{\label{tbl2}Secondary SC order parameters. The
primary SC order parameters are listed in the first column
and all secondary SC order parameters are listed in the
second column.}   
\begin{ruledtabular}
\begin{tabular}{ccc}
Primary &  Secondary  \\
\hline
$(1)$ & none \\
\hline
$(1,0)$ & none \\
$(\phi_1,\phi_2)$ & $(1)$ \\
$(\eta_1,\eta_2)$ & $(1)$ \\
\hline
$(1,0,0)$ & none \\
$(1,1,1)$ & $(1)$ \\
$(1,\eps,\eps^2)$ & $(1,0)$\\
$(|\eta_1|,i|\eta_2|,0)$ & none\\
$(|\eta_1|,|\eta_2|,0)$ & none\\
$(\eta_1,\eta_2,0)$ & none\\
$(|\eta_1|,i|\eta_2|,|\eta_3|)$ & $(1)$, $(\phi_1,\phi_2)$\\
$(|\eta_1|,|\eta_2|,|\eta_3|)$ &  $(1)$, $(\phi_1,\phi_2)$\\
$(\eta_1,\eta_2,\eta_3)$ &  $(1)$, $(\eta_1,\eta_2)$
\label{second}
\end{tabular}
\end{ruledtabular}
\end{table}

\section{Strains and elastic moduli}
Unconventional SC states normally break spatial symmetry in addition to gauge.
If the crystallographic class changes, one can expect the development 
of new components of the strain tensor and certain anomalies in the elastic moduli which can be 
measured by ultrasound propagation.\cite{Bruls90} Such a measurement has not yet been reported for
\pr. Thus, here we consider all representations for the normal-to-A phase transition.

The elastic energy for $T_h$ is the same as for $O_h$,
\begin{eqnarray}
F_{el} & =& \frac{C^0_{11}}{2}(e_1^2 + e_2^2 + e_3^2)
 + C^0_{12}(e_1e_2+e_2e_3+e_1e_3) \nonumber \\
& & + \frac{C_{44}^0}{2}(e_4^2+e_5^2+e_6^2),
\end{eqnarray}
where $e_{1,\ldots,6}$ are the components of the strain.
Generally, if the strain is a secondary order parameter it couples to
the primary order parameter as $\eta^2 e$, which leads to
a development of the secondary order parameter as $e\sim \eta^2$.

The development of the strains following each normal-to-SC
transition and discontinuities of the elastic moduli
are shown in Table~\ref{tbl3}.   

\subsection{1D order parameter}
There is no difference between $O_h$ and $T_h$ in this case.
The coupling of the strain to the SC order parameter is described
by the following term in the Landau potential
\be
F_{\eta e} = \rho|\eta|^2(e_1+e_2+e_3)
\ee
The dilatation al strain $e_1+e_2+e_3$ appears as a secondary order parameter,
and the only elastic constant which is discontinuous is $C_{11}$.

\subsection{2D order parameter}
The coupling terms are
\begin{eqnarray}
F_{\eta e} &=& \rho_1(|\eta_1|^2 + |\eta_2|^2)(e_1+e_2+e_3) \nonumber \\
& & + \rho_2[\eta_1\eta_2^{*}(e_1+\eps e_2+\eps^2 e_3) + c.c.]
\nonumber \\
& & + i\rho_3[\eta_1\eta_2^{*}(e_1+\eps e_2+\eps^2 e_3) - c.c.].
\end{eqnarray}
The third term is absent in $O_h$. The free energy of the OP is given by Eq.~(\ref{F_Eg}), which 
describes the second order phase transitions between the normal state and the superconducting 
states $(1,0)$ and $(\phi_1,\phi_2)$.

Diatomic strains $e_2-e_3$ and $2e_1-e_2-e_3$ appear in the transition to
$(\phi_1,\phi_2)$. Therefore, it is necessary to average the elastic moduli in all three 
directions to take into account domains.

\subsection{3D order parameter}
The  coupling terms are
\begin{eqnarray}
F_{\eta e} & = & \rho_1(|\eta_1|^2+|\eta_2|^2+|\eta_3|^2)(e_1+e_2+e_3)\nonumber\\
& & +\rho_2[3(|\eta_2|^2-|\eta_3|^2)(e_2-e_3)\nonumber \\
& & \quad +(2|\eta_1|^2-|\eta_2|^2-|\eta_3|^2)(2e_1-e_2-e_3)] \nonumber\\
& & + \rho_3[(|\eta_2|^2-|\eta_3|^2)(2e_1-e_2-e_3)\nonumber \\
& & \quad -(2|\eta_1|^2-|\eta_2|^2-|\eta_3|^2)(e_2-e_3)] \nonumber \\
& & +\rho_4[(\eta_2^{*}\eta_3+\eta_2\eta_3^{*})e_4
  +(\eta_3^{*}\eta_1+\eta_3\eta_1^{*})e_5 \nonumber \\
& & \hspace{.2in} +(\eta_1^{*}\eta_2+\eta_1\eta_2^{*})e_6].
\end{eqnarray}
The third term appears in $T_h$ but not $O_h$.   Shear strains $e_{4,5,6}$,
but not diatomic strains,
are present when all three components of the OP have the same magnitude.
Diatomic strains appear when any of the magnitudes differ.  

\begin{table*}[t]
\caption{\label{tbl3} Strains and discontinuities in
the elastic moduli following normal-to-SC phase
transitions in $T_h$ crystals.  The SC states are listed in the first column.
Strains which appear as secondary order parameters and discontinuities of
the elastic moduli are listed in the second and third columns respectively,
as functions of the primary order parameter and the phenomenological constants. 
The fourth-order coefficients $\beta_i$ in the Landau potential for the 2D order parameter a 
defined
in Eq.~(\ref{F_Eg}). For the 1D and 3D order parameter they correspond to the following 
terms,\cite{Sigrist1991} $\beta |\eta|^4$ and
$\beta_1(|\eta_1|^2+|\eta_2|^2+|\eta_3|^2)^2 + \beta_2 |\eta_1^2+\eta_2^2+\eta_3^2|^2 +
\beta_3 (|\eta_1|^2|\eta_2|^2+|\eta_2|^2|\eta_3|^2+|\eta_1|^2|\eta_3|^2)$, respectively.
The domain average values for the elastic moduli $C_{ij}$ are calculated 
as $C_{11}^{\text{Av}}=(C_{11}+C_{22}+C_{33})/3$, $C_{12}^{\text{Av}}=(C_{12}+C_{23}+C_{13})/3$. 
The superscript
0 denotes the values in the normal state.}
\begin{ruledtabular}
\begin{tabular}{cll}
Transition: &  Strains which appear & Elastic moduli \\
Normal to & as secondary order parameters & in the SC state\\
\hline
$(1)$ & $e_1+e_2+e_3  = \frac{-3\rho|\eta|^2}{C_{11}^0+2C_{12}^0}$ &
$C_{11} = C_{22} = C_{33} = C_{11}^0-\frac{\rho^2}{2\beta}$\\
& & $C_{11}-C_{12}$, $C_{44}$ continuous \\
\hline
$(1,0)$ &
$e_1+e_2+e_3 = \frac{- 3\rho_1|\eta|^2}{C_{11}^0+2C_{12}^0}$ &
$C_{11} = C_{22} = C_{33} = C_{11}^0 - \frac{\rho_1^2}{2\beta_1}$ \\
& & $C_{11}-C_{12}$, $C_{44}$ continuous\\
\hline
&$e_1+e_2+e_3 = \frac{-6\rho_1|\eta_1|^2}{C_{11}^0+2C_{12}^0}$ &
$C_{11}^\text{Av}  =  C_{11}^0 - \frac{2\rho_1^2+\rho_2^2+\rho_3^2}
{2\beta_1+\beta_2}$ \\
$(\phi_1,\phi_2)$ & $2e_1-e_2-e_3  = \frac{-6|\eta_1|^2(\rho_2\cos\phi-\rho_3\sin\phi)}
{C_{11}^0-C_{12}^0}$ &
$C_{12}^\text{Av}  =  C_{12}^{0} - \frac{4\rho_1^2-\rho_2^2
-\rho_3^2}{2(2\beta_1+\beta_2)}$
 \\
& $e_2-e_3  = \frac{2\sqrt{3}|\eta_1|^2(\rho_2\sin\phi+\rho_3\cos\phi)}
{C_{11}^0-C_{12}^0}$ &  $C_{44}$ continuous\\
\hline
&$e_1+e_2+e_3 = \frac{-3\rho_1|\eta_1|^2}{C_{11}^0+2C_{12}^0}$ &
$C_{11}^\text{Av}  =  C_{11}^0 -\frac{3\rho_1^2+24\rho_2^2+8\rho_3^2}{6(\beta_1+\beta_2)}$\\
$(1,0,0)$ &
$2e_1-e_2-e_3 = \frac{-12\rho_2|\eta_1|^2}{C_{11}^0-C_{12}^0}$&
$C_{12}^\text{Av}  =  C_{12}^0 - \frac{3\rho_1^2-12\rho_2^2-4\rho_3^2}
{6(\beta_1+\beta_2)}$\\
&$e_2-e_3 = \frac{4\rho_3|\eta_1|^2}{C_{11}^0-C_{12}^0}$
& $C_{44}$ continuous\\
\hline
&$e_1+e_2+e_3 = \frac{-9\rho_1|\eta_1|^2}{C_{11}^0+2C_{12}^0}$ &
$C_{11} = C_{22} = C_{33} = C_{11}^0 - \frac{3\rho_1^2}{2(3\beta_1+3\beta_2+\beta_3)}$ \\
$(1,1,1)$ &
$e_{4,5,6} = - \frac{2\rho_4|\eta_1|^2}{C_{44}^0}$ &$C_{11}-C_{12}$ continuous\\
&
& $C_{44} = C_{44}^0 - \frac{2\rho_4^2}{3(3\beta_1+3\beta_2+\beta_3)}$\\
\hline
& $e_1+e_2+e_3 = \frac{-9\rho_1|\eta_1|^2}
{C_{11}^0+2C_{12}^0}$ & $C_{11} = C_{22} = C_{33} = C_{11}^0-\frac{3\rho_1^2}{2(3\beta_1+\beta_3)}$ \\
$(1,\eps,\eps^2)$   & $e_{4,5,6} = \frac{\rho_4|\eta_1|^2}{C_{44}^0}$
& $C_{11}-C_{12}$ continuous \\
& & $C_{44} = C_{44}^0 -\frac{\rho_4^2}{6(3\beta_1+\beta_3)}$\\
\hline
& $e_1+e_2+e_3 = \frac{-3\rho_1(\eta_1^2+\eta_2^2)}{C_{11}^0+2C_{12}^0}$&
$C_{11}^\text{Av} = C_{11}^0 -\frac{3\rho_1^2(4\beta_2-\beta_3)+4(3\rho_2^2+\rho_3^2)
(6\beta_1+2\beta_2+\beta_3)}{6(4\beta_2-\beta_3)(4\beta_1+\beta_3)}$
\\
$(|\eta_1|,i|\eta_2|,0)$ &
$e_1+e_2-2e_3 = \frac{-6\rho_2(|\eta_1|^2+|\eta_2|^2)+6\rho_3(|\eta_1|^2-|\eta_2|^2)}{C_{11}^0-C_{12}^0}$&
$C_{12}^\text{Av} = C^0_{12}-\frac{3\rho_1^2(4\beta_2-\beta_3)-2(3\rho_2^2+\rho_3^2)
(6\beta_1+2\beta_2+\beta_3)}{6(4\beta_2-\beta_3)(4\beta_1+\beta_3)}$ \\
& $e_1-e_2 = \frac{-6\rho_2(|\eta_1|^2-|\eta_2|^2)-2\rho_3(|\eta_1|^2+|\eta_2|^2)}{C_{11}^0-C_{12}^0}$ & $C_{44}$ continuous
\end{tabular}
\end{ruledtabular}
\end{table*}

\section{Discussion}

Experimentally, the symmetry of the SC states and the
nature of the phase transition between them in \pr\ is  far from resolved.
Anomalies at $T_{c2}$ have been observed in many
experiments~\cite{Vollmer2003,Oeschler2003,Aoki2002,Aoki2003,Ho03,Broun,Izawa2003,Tayama03,Chia03}.
Specific heat measurements by
Volume {\em et al.}\cite{Vollmer2003}  found a jump at $T_{c2}$, indicative of a second-order
phase transition.  On the other hand, Aoki {\em et al.}\cite{Aoki2002,Aoki2003} found a kink, 
resulting in a steeper temperature dependence below $T_{c2}$, which seems to  correspond to a
first-order phase transition.
The most dramatic observation is the change in symmetry at the A-B phase
transition seen in thermal conductivity measurements.\cite{Izawa2003}
The double transition was also observed in magnetisation measurements as a peak effect in 
$M(H)$.\cite{Tayama03,Ho03}
One of these measurements found
strong anisotropies,\cite{Tayama03} possibly indicative of a change in
symmetry;  the other did not.\cite{Ho03}
Finally, recently penetration depth measurements have been interpreted not
as a phase transition, but rather as a cross-over due to two-band
superconductivity.\cite{Broun}

The temperature range in which the A-phase exists is very narrow, thus with two
exceptions,\cite{Izawa2003,Ho03} the reported experiments probe the properties
of the gap in the B-phase. Experiments consistently rule out the existence of line nodes in the 
B-phase.\cite{Bauer2002,Kotegawa03,Macl02,Izawa2003,Suderow03}
However, the presence of point nodes in the B-phase is clearly indicated
by a power law temperature dependence of the specific
heat,\cite{Bauer2002} the thermal conductivity measurement,\cite{Izawa2003}
and the penetration depth.\cite{Chia03}
Nuclear quadrupolar resonance experiments\cite{Kotegawa03} can be interpreted as either
fully gapped or nodes. Very low temperature tunneling spectroscopy~\cite{Suderow03}
finds no nodes at all in the B-phase, perhaps consistent with rigorous nodes,
rather than approximate nodes. Finally,  $\mu$SR\cite{Macl02} indicates that the B-phase 
is fully gapped.

Only a couple of experiments have specifically dealt with the symmetry of the
gap function.\cite{Izawa2003,Chia03} In the thermal conductivity experiment,
point nodes were found in the [010]
direction in the B-phase
and in both the [100] and [010]  directions in the
A-phase.\cite{Izawa2003}   However, in this measurement,
there is no clear
explanation for why the two-fold symmetry is actually observed as such, rather than averaged out
into domains.
The penetration depth has a power law temperature dependence
corresponding to
point nodes along all three principal crystallographic axes.\cite{Chia03}
No studies of
the nodal structure along the $[111]$ direction have been reported so far.
An extremely important finding is due to another $\mu$SR measurement, which showed that
time reversal symmetry is broken in the B-phase.\cite{Aoki2003}

In determining which of the states listed in Table I best describes
\pr\ we make the following assumptions:
{\em i}) the B-phase breaks time reversal symmetry;
{\em ii}) there are point nodes in the B-phase located in the [100]
and/or equivalent directions, and there are no line nodes in the B-phase;
{\em iii}) the A-B phase transition is second order;
{\em iv}) both phases are described by the same order parameter.
The first two assumptions are based on fairly conservative
interpretations of the experimental data available to date.
We use the last two assumptions to narrow the choices of possible states.
Their validity is subject to further experimental study.

We exclude the $A$ and $E$  representations because of {\em ii}).
In $T_g$ and $T_u$ representations, the
first four states listed in Table~\ref{tbl}
are
connected to the normal state by a second-order phase transition
(see Fig.~\ref{fig2}), but among them only $(1, 0, 0)$ and $(|\eta_1|, i|\eta_2|,0)$ may
be followed by another second order phase transition involving the
same order parameter.   Therefore,
these are the only two possibilities for the A-phase.
If the A-phase is $(|\eta_1|, i|\eta_2|, 0)$ then the
B-phase is either $(\eta_1, \eta_2, 0)$ or $(|\eta_1|, i|\eta_2|, |\eta_3)$.
The former is excluded because it has line nodes in the singlet channel and
no nodes at all in the triplet channel. The latter possibility must be
singlet because it has no nodes at all in the triplet channel.
If the A-phase is $(1, 0, 0)$ then the B-phase is the $(|\eta_1|, i|\eta_2|, 0)$ state.
Because there are no line nodes in
the B-phase, the pairing is therefore triplet.  Strictly speaking,
 $(|\eta_1|, i|\eta_2|, 0)$ has no nodes at all under $T_h$ symmetry.
However, nodes appear in the corresponding $O_h$  state $(1,i,0)$.\cite{Volovik1985}
Such nodes may be pronounced dips in $T_h$ if the Fermi surface has the approximate $O_h$ symmetry
as found in Ref.~\onlinecite{Sugawara02}. Therefore, the two most likely possibilities for the 
sequence of SC phase transitions in \pr\ are
$$
\text{normal} \rightarrow (|\eta_1|,i|\eta_2|,0) \rightarrow
(|\eta_1|, i|\eta_2|, |\eta_3|)
$$
in the singlet channel and
$$
\text{normal} \rightarrow (1,0,0)\rightarrow  (|\eta_1|,i|\eta_2|,0)
$$
in the triplet channel.

If future experiments fail to be consistently described  within the framework
described in this article, then 
it is likely that the 
assumption that the order parameters of both transitions
belong to
the same representation 
will merit closer examination.  
It is possible that the B-phase may be due to the appearance
of an order parameter that belongs to a different representation than that
of the A-phase.
This possibility is somewhat unsatisfactory
in situations when the phase transitions occur very close 
together, as in \pr, because it suggests a
rather fine tuning of the phenomenological parameters.
Second-order phase transitions between any states which are related 
as group-subgroups are allowed, provided third order terms of the effective
order parameter are absent in the free energy.  
The order parameter of the B-phase may be
a superconducting order parameter that belongs to a different representation than that of the
A-phase, or it could even be something completely different, such as a
structural order parameter, or a state with broken translational symmetry.

\section{Summary}

To summarise, we find group-theoretically the SC states which can be realized in
crystals with $T_h$ symmetry. Additional symmetry breaking within the
SC state is considered. Heavy fermion superconductivity in \pr\ is best described by the
three-dimensional representations of $T_h$ point group. Considering experimental
results, we
propose the two most likely scenarios for the SC phase transition sequence found in \pr, one
in the singlet and another in the triplet channel.

\begin{acknowledgments}
We greatly appreciate discussions with Y. Aoki, H. Harima, I. Vekhter, K. Ueda, and S. Urazhdin.
This work was supported by NSERC of Canada.
\end{acknowledgments}


\begin{thebibliography}{99}
\bibitem{Maple2001} M. B. Maple, E. D. Bauer, V. S. Zapf, E. J. Freeman,
N. A. Frederick and R. P. Dickey, \Journal{Acta Phys. Pol. B}{32}{3291}{2001}.

\bibitem{Bauer2002} E. D. Bauer,  N. A. Frederick, P.-C. Ho, V. S. Zapf, and M. B. Maple,
\Journal{\PRev\ B}{65}{100506(R)}{2002}.

\bibitem{Kotegawa03} H. Kotegawa, M. Yogi, Y. Imamura, Y. Kawasaki, G.-q. Zheng, Y. Kitaoka,
S. Ohsaki, H. Sugawara, Y. Aoki, and H. Sato, \Journal{\PRev\ Lett.}{90}{027001}{2003}.

\bibitem{Aoki2002} Y. Aoki, T. Namiki, S. Ohsaki, S. R. Saha, H. Sugawara and H. Sato, 
\Journal{J. Phys. Soc. Jpn.}{71}{2098}{2002}.

\bibitem{Macl02} D. E. MacLaughlin, J. E. Sonier, R. H. Heffner, O. O. Bernal, B.-L. Young,
M. S. Rose, G. D. Morris, E. D. Bauer, T. D. Do, and M. B. Maple, 
\Journal{\PRev\ Lett.}{89}{157001}{2002}.

\bibitem{Vollmer2003} R. Vollmer, A. Fai$\beta$t, C. Pfleiderer, H. v. L\"ohneysen, E. D.
Bauer, P.-C. Ho, V. Zapf, and M. B. Maple, \Journal{\PRev\ Lett.}{90}{057001}{2003}.

\bibitem{Oeschler2003} N. Oeschler, P. Gegenwart, F. Steglich, N. A. Frederick,
E. D. Bauer and M. B. Maple, \Journal{Acta Phys. Polon. B}{34}{959}{2003}.

\bibitem{Izawa2003} K. Izawa, Y. Nakajima, J. Goryo, Y. Matsuda, S. Osaki, H. Sugawara,
H. Sato, P. Thalmeier, and K. Maki, \Journal{\PRev\ Lett.}{90}{117001}{2003}.

\bibitem{Sugawara02} H. Sugawara, S. Osaki, S. R. Saha, Y. Aoki, H. Sato, Y. Inada, 
H. Shishido, R. Settai, Y. \=Onuki, H. Harima, and K. Oikawa, \Journal{Phys. Rev. B}
{66}{220504(R)}{2002}.

\bibitem{Kohgi03} M. Kohgi, K. Iwasa, M. Makajima, N. Metoki, S. Araki, N. Bernhoeft,
J.-M. Mignot, A. Gukasov, H. Sato, Y. Aoki and H. Sugawara, \Journal{\JPSJ}{72}{1002}{2003}.

\bibitem{Cao03} D. Cao, F. Bridges, S. Bushart, E. D. Bauer, and M. B. Maple, 
\Journal{\PRev\ B}{67}{180511(R)}{2003}.

\bibitem{Tayama03} T. Tayama, T. Sakakibara, H. Sugawara, Y. Aoki and H. Sato, 
\Journal{\JPSJ}{72}{1516}{2003}.

\bibitem{Ho03} P.-C. Ho, N. A. Frederick, V. S. Zapf, E. D. Bauer, T. D. Do, M. B. Maple,
A. D. Christianson, and A. H. Lacerda, \Journal{\PRev\ B}{67}{180508(R)}{2003}.

\bibitem{Frederick03a} N. A. Frederick and M. B. Maple, 
\Journal{J. Phys.: Condens. Matter}{15}{4789}{2003}.

\bibitem{Aoki2003} Y. Aoki, A. Tsuchiya, T. Kanayama, S. R. Saha, H. Sugawara, H. Sato, 
W. Higemoto, A. Koda, K. Ohishi, K. Nishiyama, and R. Kadono, 
\Journal{\PRev\ Lett.}{91}{067003}{2003}.

\bibitem{Suderow03} H. Suderow, S. Viera, J. D. Strand, S. Bud'ko and P. C. Canfield,
cond-mat/0306463.

\bibitem{Frederick03b} N. A. Frederick, T. D. Do, P.-C. Ho, N. P. Butch, V. S. Zapf, and 
M. B. Maple, cond-mat/0307059.

\bibitem{Chia03} E. E. M. Chia, M. B. Salamon, H. Sugawara, and H. Sato,
\Journal{\PRev\ Lett.}{91}{247003}{2003}. 

\bibitem{Broun} D. M. Broun, P. J. Turner, G. K. Mullims, D. E. Sheehy, X. G. Zheng,
S. K. Kim, N. A. Frederick, M. B. Maple, W. N. Hardy, and D. A. Bonn, cond-mat/0310613.

\bibitem{Anders2002} F. B. Anders, \Journal{Eur. Phys. J. B}{28}{9}{2002}.

\bibitem{Goryo2003} J. Goryo, \Journal{\PRev\ B}{67}{184511}{2003}.

\bibitem{Maki2003} K. Maki, H. Won, P. Thalmeier, Q. Yuan, K. Izawa and Y. Matsuda,
\Journal{Europhys. Lett.}{64}{496}{2003}.

\bibitem{Ichioka2003} M. Ichioka, N. Nakai, and K. Machida, \Journal{\JPSJ}{72}{1322}{2003}.

\bibitem{Miyake2003} K. Miyake, H. Kondo, and H. Harima, 
\Journal{J. Phys.: Condens. Matter.}{15}{L275}{2003}.

\bibitem{Asano03} Y. Asano, Y. Tanaka, Y. Matsuda, and S. Kashiwaya, 
\Journal{\PRev\ B}{68}{184506}{2003}.

\bibitem{Volovik1985} G. E. Volovik and L. P. Gor'kov, \Journal{Sov. Phys. JETP}{61}{843}{1985}.

\bibitem{Landau68} L. D. Landau, E. M. Lifshitz, \emph{Statistical Physics}, pt.1, 
3rd rev. and enl. ed., Pergamon Press, New York, 1978.

\bibitem{Sigrist1991} M. Sigrist and K. Ueda, \Journal{Rev. Mod. Phys.}{63}{239}{1991}.

\bibitem{HarimaNote} Although $4f$ electrons of Pr in PrOs$_4$Sb$_{12}$ are 
localized,\cite{Sugawara02} spin-orbit coupling strongly affects the Fermi surface
due to $5d$ electrons of Os and $5p$ electrons of Sb (H. Harima, private 
communication).

\bibitem{Gufan1995}  Yu. M. Gufan, \Journal{Sov. Phys. JETP}{80}{485}{1995}.

\bibitem{mistake} We also fix several errors found in Ref.~\onlinecite{Gufan1995}.

\bibitem{Tinkham} M. Tinkham, \emph{Group theory and quantum mechanics}, New York, 
McGraw-Hill, 1964.

\bibitem{TediousNote} For example, it can be shown that in order to describe two consequtive
second order phase transitions described by a multidimensional order parameter one needs to 
expand the Landau potential up to eigth order terms at least.

\bibitem{Blount1985} E. I. Blount, \Journal{\PRev\ B}{32}{2935}{1985}.

\bibitem{Yip93} S. Yip and A. Garg, \Journal{\PRev\ B}{48}{3304}{1993}.

\bibitem{Monien1986} H. Monien, K. Scharnberg, L. Tewordt, and N. Schopohl, 
\Journal{\PRev\ B}{34}{3487}{1986}; \Journal{J. Low Temp. Phys.}{65}{13}{1986}.

\bibitem{ComNodes} Function~(\ref{gapEg}) has accidental line nodes.

\bibitem{Bruls90} G. Bruls, D. Weber, B. Wolf, P. Thalmeier, B. L\"uthi, A. de Visser, and 
A. Menovsky, \Journal{Phys. Rev. Lett.}{65}{2294}{1990}.

%\bibitem{Slonczewski70} J. C. Slonczewski and H. Tomas, \Journal{\PRev\ B}{1}{3599}{1970}.

\end{thebibliography}
\end{document}